\documentclass{article}
\usepackage{epsfig}
\title{Study of Spectral Lag  for Short GRBs }
\author{Varsha Gupta,\thanks{E--mail : varsha@physics.du.ac.in}
          Patrick Das Gupta\thanks{E--mail : patrick@ducos.ernet.in}\\
    {\em Department of Physics and Astrophysics,} \\
     {\em University of Delhi, Delhi-110 007, India.}\\
          and P.N. Bhat\thanks{E--mail : pnbhat@tifr.res.in}\\
          {\em Tata Institute of Fundamental Research,}\\
          {\em Homi Bhabha Road, Mumbai-400 005, India.}}

\begin{document}
%\singlespacing
\maketitle
\large
\newcommand{\del}{\mbox{$\partial$}}
\begin{abstract}
This paper reports statistically significant correlations between
various burst
 parameters, observed in a sample of 156 GRBs belonging to BATSE 4B catalog
 with T90 less than 2 s. The number of subpulses in a burst is strongly
 correlated not only with the object's duration but also with its fluence and
 hardness ratio, suggesting that when the central engine is
  more powerful, ejecting matter with typically higher
 values of Lorentz factor, the bulk energy is dissipated on longer time scales
 in the form of
 larger number of gamma pulses. We estimate hard-to-soft lag in bursts by
 taking the difference between centroids corresponding to time profiles at
 energies $> 100$ keV and $<100$ keV.
 The number of short GRBs that show soft-to-hard spectral evolution is
  slightly over one quarter  of the total, in the sample considered here.

 Bursts that exhibit hard-to-soft spectral change appear to form a distinct
 class, with strength as well as hardness of individual subpeaks
 tending to decrease with peak position. Opposite is true for objects with
 softer photons
 arriving earlier than the harder ones, implying some kind of a rejuvenation of
 the central engine (may be due to enhanced accretion of matter towards the end). The two classes
 also show other diverging trends. For instance, objects belonging to the larger of the two classes display
 strong correlations between spectral lag and the fluence, the hardness ratio
 as well as the number of pulse, respectively. While no such correlations are seen in
 bursts that evolve from soft to hard. However, the magnitude of lag is
 strongly correlated with burst duration in both the classes.
\end{abstract}
\begin {section}{Introduction}

More than circumstantial evidence point towards short GRBs
 forming a  distinct subclass of bursts. Bursts with T90 less than $\sim$
 2 s are harder [1]
 and their fluences are about $\sim $ 20 times weaker than their longer
  counterparts [2,3]. From the lag obtained by cross correlating  TTE
 data corresponding to 25-50 KeV and 100-300 KeV channels, Norris, Scargle
 \& Bonnell found
 that GRBs with duration $<$ 2.6 s not only exhibit about an order of magnitude lower values of
 lag than those of long bursts but also these values appear to be discontinuous
across $\sim $ 2 s [4]. Another
 study by  Paciesas
 et al. seems to indicate that the average spectral break energy for short
  bursts is larger than that of long duration GRBs by a factor of about $\sim $
 2 signifying for the former either larger bulk Lorentz factors or spatially
 closer location [5].

 However, while studying various pulse
 parameters for a sample of 100 bright short GRBs, McBreen et al. recently
 concluded that most of the strong correlations observed are similar to those
 found in the case of long bursts suggesting identical gamma-emission
 mechanism for both the classes [6]. Absence of directly measured redshifts due
 to  lack of observed optical afterglows in the case of short bursts makes it
 difficult to verify the
 lag-luminiosity relation as seen by Norris, Marani \& Bonnell in some of the long duration GRBs [7]. Evidently, a lot of
work still needs to be carried out in order to properly understand short bursts.

In this paper, we focus on a sample of short bursts belonging to the BATSE
4B catalog. Individual peaks in a burst are identified using a zero crossing
algorithm developed by Bhat et al. [8].
  A simple measure to characterize  lag between
 hard and soft gamma rays is considered in the present work  to study
 correlations among various burst parameters.
\end{section}
\begin {section}{Data set and `hard-soft' time lag}

 We begin with all the category A bursts (defined to be those lying
completely within the TTE window) listed in the BATSE 4B catalog with T90 less than
 2 s [9]. There are about 182 such objects. Since we require post-burst background
with duration equal to that of the actual GRB, for
 identifying  individual pulses in a burst [8,10], the final sample that is
  subjected
to various analysis in this paper consists of 156 objects. For the purpose of
peak detection in a burst, TTE data for all the 4 energy channels are combined. Fluence,
 hardness ratio and T90 along with the respective errors are taken from
the available BATSE catalog itself.

If $t_i$ represents the arrival time of $i^{th}$ photon in the detector then
the time-centroid is simply expressed as,
\begin{equation}
t_{cen}=\frac {1}{N} \sum^k_j t_i
\end{equation}
where $N$ is the number of photons detected between the starting
 ($t_j$) and the end ($t_k$) instants of the burst.

Spectral lag, in GRB literature, is usually estimated by cross-correlating
time histories of a burst determined at distinct energy bands (e.g.
[11,4]).
In the present work, we use a simple definition of `hard-soft' time
 lag $\tau_c$ for a given burst.
 This lag is computed by taking the difference between
 time-centroids obtained from arrival times corresponding to photon energies
larger than 100 keV and from those  less than 100 KeV, so that
\begin{equation}
\tau_c \equiv (t_{cen})_{3+4} - (t_{cen})_{1+2},
\end{equation}
where the subscripts 1,2,3 and 4 represent the four distinct BATSE energy
 channels.
 This
 simple characterization of lag is similar to a  definition of
  energy evolution
  parameter used by Bhat et
al.[8]. A negative value of $\tau_c $ correspond to harder photons arriving
 earlier than those  with energy $< $ 100 keV, and vice versa. As the
 photon arrival times are recorded with a high resolution of 2 $\mu $s,
 the statistical error in the determination of $t_{cen}$ is given by,
\begin{equation}
\Delta t_{cen} = \frac {2\ \mu s}{\sqrt N},
\end{equation}
which typically is 2 orders of magnitude lower than the arrival time
 uncertainty.

We find 116 short bursts with negative $\tau_c$ ranging from $-0.05 \pm 4 \times 10^{-8} $s
 (for GRB trigger no.3215) to $-7.27 \times 10^{-5} \pm 1 \times 10^{-7} $s (for  trigg.no.2395). While
 rest of the 40 objects
 exhibit positive values, indicating a soft-to-hard evolution, lying in
 the interval ($9 \times 10^{-5} \pm 6 \times 10^{-8} $s, $0.03 \pm 4\times 10^{-8} $s) with the
 boundary points corresponding to trigger no.s 1308 and 1791, respectively.
 A histogram of $\tau_c$ for all the 156 bursts is shown in Fig.1.
 The gross features of the distribution are similar to the ones seen by Norris
 et al. [4]. Unlike in the case of long bursts where one finds a small fraction
  (less than $\sim $ 0.1) of sources displaying soft-to-hard evolution [11],
 the percentage of
 short GRBs
 with positive $\tau_c$ is $\sim  26 \% $, which is larger by about a factor
 of three.
  Since a substantial number of
 short bursts appear to undergo soft-to-hard spectral change,
 we also present many of the significant
 statistical correlations
  in the following section separately for
 $\tau_c $ $> 0$ and $< 0$ cases, respectively.
\end{section}
\begin {section}{Burst Parameters and Linear Correlations}
Using the peak detection method (for details see [10]), the subpulses present
in a burst are identified at an optimum resolution (defined to be the time
resolution at which the algorithm returns the largest  number of peaks for that burst). Once this is done, the peak position relative to the starting instant of the burst as well as the peak height for each subpulse can be determined. The
total number of peaks found in a given burst has been termed as its complexity index (CI) in this paper.

In our sample, we find that the average value of CI is between
2 and 3, with most bursts having fewer than 5 subpulses. There are 45
 single peaked GRBs, while only about $12 \%$ of the total have CI lying in the range 5 to a
maximum value of 8. In contrast, long duration bursts tend to be much more
complex with CI usually exceeding 10 [12,13]. Among the short bursts, for both negative as well as positive
 $\tau_c$ cases, we find that CI is strongly correlated each with T90, fluence
  and
 hardness ratio (HR).
  Similarly, T90 displays significant correlation  with $\vert { \tau_c} \vert $ as
well as fluence over the entire sample. For all the 156 short bursts, there
 is also a strong correlation between fluence and HR. The results are listed
 in table 1, where the correlation coefficients have been computed using
  the weighted linear correlation analysis (see e.g.[14]).

  Weighted correlation
 incorporates error $\sigma_i$ for each data pair ($x_i$,$y_i$),
\begin {equation}
\sigma_i= \sqrt {\sigma^2_{xi} + \sigma^2_{yi} }
\end {equation}
where $\sigma_{xi}$ and $\sigma_{yi}$ are the errors on $x_i$ and $y_i$,
respectively. For CI, we have set the error to be zero. Whenever variables
 carry different physical dimensions (like T90 and fluence,
for instance), in order that eq.(4) remains meaningful we redefine the
 quantities by taking ratios of
$x_i$ and $y_i$ with $x_{max}$ and $y_{max}$, respectively, where $x_{max}$
     and $y_{max}$ are the corresponding maximum values in the sample.
The redefined variables along with the respective errors are dimensionless, and
 are therefore amenable to weighted correlation analysis.

% For the 156 objects, the
% weighted correlation coefficient corresponding to the pair CI and T90, with
% latter's error taken into account, is 0.58. The probability for such a
% high degree of correlation to be
% purely due to chance is $     $. Similarly, with errors on fluence
% included, the weighted correlation coefficient for CI and fluence
% and the corresponding chance probability are 0.43 and $  $, respectively.
  For the sample of 156 objects studied in this paper, though
 T90 information is available
 for each, the 4B BATSE catalog lists the measured values of fluence
 and HR for 149 (of which 37 burst have positive $\tau_c$) and 133 (33 objects
of these show soft-to-hard evolution) bursts, respectively. These numbers
 have been used in tables 1 and 2 to compute chance probabilities to check
 for significant
 correlations.

 It may be appropriate at this juncture to speculate on a general picture
 of short GRBs that the correlations in table 1  appear to be throwing light on.
 A central engine which pumps energy into
 the ambient medium at a variable rate followed by internal shocks developing
 due to collision of multiple winds is presently the most promising theoretical
  scenario to describe GRB phenomena [15,16].
 In view of such models, the observed strong
 correlation between CI and T90  appears to suggest that
 when the central engine is active for a longer
 duration, the resulting gamma emission is often in the form of several
 subpulses
 instead of taking the shape of a broad single peak. Therefore, the
 result seems to favour a multiple collisions of wind-like ejecta
 scenario for longer short bursts.

 One needs to exercise caution regarding
 the correlation  we observe between T90 and fluence, mainly
 because of two factors.
  Firstly, although for the rest of the pairs of burst parameters, the correlation coefficients are
 large  as well as consistent with results presented in table 1,
  when one carries out a
 linear correlation analysis without errors, there is a difference
 in the case of T90-fluence pair. If one considers a linear correlation
analysis for this pair without including errors then the coefficient is only 0.05, which is statistically insignificant.
 However,
according to an older study on a sample of 400 GRBs by Lee and Petrosian, longer  bursts statistically tend to of higher fluence [17]. So one is tempted to
 accept the
 strong correlation we observe in our analysis.

 But then, one needs to be careful as another factor comes into the picture. It has been recently pointed out by Hakkila et al.
  that  fluence-duration bias in
 the BATSE data may cause underestimation of fluence as well as duration for
 faint GRBs [18]. However, this bias may not cause
  high fluence, long duration GRBs
  to appear as low fluence, short and hard bursts [19]. In a sense, it is unlikely that short bursts as a class may get contaminated due to this bias.

 Keeping the above discussion in mind, we may proceed  with our analysis.
It is evident from table 1 that harder the burst,  more
 complex it is. Also, higher the fluence, harder is the burst. In a way these
 trends are natural, since if the amount of energy released by the central
 engine  is larger then Lorentz
 factors as well as degree of activity and complex interactions
  between multiple
 winds are expected to be higher.
  Similarly, T90 displays a strong correlation with $\vert { \tau_c} \vert$ over the entire sample. We will touch upon this point again when we discuss
positive and
negative $\tau_c$ cases as distinct subclasses.

In table 2, we present correlation coefficients separately for bursts showing
 soft-to-hard
 and hard-to-soft spectral changes. What is striking is that  for negative
  $\tau_c $ cases (we define them to be Class A bursts, not to be confused
 with category A objects of BATSE data [9]),
  the spectral lag shows a strong anti-correlation with CI, HR as well as
  fluence.  In
   contrast, no significant correlation is exhibited by the corresponding
 quantities  for bursts with $\tau_c > 0$ (termed as Class B objects
 henceforth).
 Hence, larger the magnitude of lag, higher is CI, HR and fluence in the case
 of Class A events, while no such trend is seen among Class B bursts.
 On the other hand, though HR and T90
  hardly appear to be correlated among bursts demonstrating
  hard-to-soft spectral progress,  this pair show a  strong
  correlation in the smaller sample of GRBs with reverse evolutionary trend.

 Now, between $\tau_c$
 and T90, there is strong statistical anti-correlation (correlation) among
 Class A (Class B) GRBs. So, what we have here is that
 irrespective of the sign of the lag, its magnitude is very strongly
 correlated with duration of bursts of A or B type.
  This basically reflects the correlation between T90
 and $\vert \tau_c \vert $ presented already in table 1.
  This feature could be
useful in constraining gamma-emission models. In general, one may argue
 that if the energy injected by the central engine
   is large then not only the burst duration is
 expected to be longer but also  typical time scale separating high and low
 velocity wind is likely to be greater.

In the next section, we will encounter further differences between Class A and
 B sources.
\end{section}
\begin {section}{Pulse Parameters versus Lag for multi-peaked GRBs}
In order to study how peak heights $p_h$ change statistically with peak
 instants $t_p$ in
 a burst, we consider a subsample of multi-peaked (CI $>1 $) short bursts
 consisting of 111 objects (of the 156 sources, 45 contain single pulses).
There are 80 multi-peaked sources with negative $\tau_c$, while  rest of the
 31 objects are of Class B type.
 It is interesting to note that only 9 out of 45 single peaked bursts are of
 Class B type.

For each burst in the subsample, we normalise the peak counts and positions
by dividing them with the maximum observed peak height and T90 of that burst,
respectively. The correlation coefficient between normalised $p_h$ and $t_p$
for Class A objects is -0.31. As there are a total of 250 pulses for 80 of these GRBs, the chance probability of finding a value of -0.31 if $p_h$ and $t_p$
 are uncorrelated is only $6.4 \times 10^{-7}$. On an average, therefore peak
 height decreases with time for bursts that show hard-to-soft evolution. When
 a similar exercise is undertaken for 31 multi-peaked Class B bursts, one finds
 a positive correlation between $p_h$ and $t_p$. The correlation coefficient
 turns out to be 0.19 corresponding to a chance probability of 0.05, since the
total number of peaks are 107. Although, the correlation is not as strong as in
the case of Class A objects, one sees a trend of increasing peak height with
 time at about 95 percent confidence level.

Of this sample of 111 bursts with multiple pulses, we now consider those
 objects with well separated peaks. A pulse is defined to be well separated
 if its height is more than double the value of photon counts of the valley on
 each side of the peak, after the background counts are subtracted from the
 time profile (also see [13]). There are 61 such bursts with all peaks well
 separated according to the above criteria. Of these, 41 short GRBs fall in
the category of Class A. Since the pulses have almost well defined beginning
  $t_b$ and
 end $t_e$ instants, determined from the position of valleys on either
 side of the peak
 maximum, we may estimate the hardness ratio of individual peaks.

 The hardness ratio ($hr_{pk}$) of a pulse is defined in this paper as the ratio
 of background subtracted photon number with energy $>$ 100 keV counted
   between the instants $t_b$ and $t_e$
  characterising the pulse, and the corresponding number
  with energy $<$ 100 keV. Total number of photons
 $n_{ph}$ for each peak is estimated by summing the counts in the interval ($t_b$, $t_e$)
 from all four energy channels. The peak flux $flx_{pk}$ corresponding to a
 pulse is simply the peak height
 $p_h$ divided by the time resolution.
  We
  quantify the symmetry $ r_{rd}$ of a pulse profile by taking the ratio,
\begin {equation}
r_{rd} \equiv \frac {tr}{td}
\end {equation}
  where $ tr$ and $td $ are the time intervals $t_p - t_b$ and $t_e- t_p$,
 respectively,  $t_p$ being
  the time instant at which peak maximum
 occurs.
 Similarly, we may also characterize the width $w$ of a peak as follows,
\begin {equation}
w \equiv \frac {t_e-t_b} {2}
 \end {equation}

 In a crude sense, $tr$ and $td$
  represent the rise and decay time scales for a pulse.
 These parameters as well as the definition of symmetry indicator $r_{rd}$,
 which rely only on measured quantities in the present work,
  somewhat differ from the ones we
 employed in our earlier study that made use of lognormal fitting
 of pulses [20].  In order to
  fit subpulses
 with, say lognormal functions, one requires starting instants of
 these peaks (see [10]), which in general are rather uncertain. These errors
 then get carried over to the parameters
 that are derived from the fits, resulting in
   large systematic errors,
  particularly for not so well
 separated peaks.
  The rational behind the present direct approach is simply to avoid such
 systematic errors.

The correlation coefficients along with chance probabilities for various
 burst parameters $X$ and $Y$ are listed in table 3. Pulse hardness
 ratio and photon number were normalised by dividing each of them by
 the respective maximum value in a given burst, while $t_p$ is
normalised by taking its ratio with burst duration.  We find from
 the table that while in the case of Class A bursts both hardness ratio
 and photon number corresponding to individual peaks tend to decrease with
 peak position, an opposite trend seems to be present in Class B objects.
It may be recollected that similar behaviour is exhibited by peak height
 for 111 multi-peaked bursts.

 Class A bursts appear to exhibit some interesting trends like
 larger the lag (i.e. more negative the $\tau_c$), harder,wider and
more asymmetric is the first peak. In the case of Class B sources, as
 $\tau_c$ increases the first peak tends to be softer and wider
 (at 90 $\%$ confidence level) while its decay time relative to the last
 peak is larger,
  although its symmetry has no significant correlation with the lag. Also, for $\tau_c >0$
 cases, if the first
 peak is harder it tends to be more symmetric.
 Whether Class A or B, ratio of $hr_{pk}$ of the first peak to last
 is strongly correlated with corresponding ratios of rise time, decay time and
 width.

For Class A objects, since peak hardness ratio is anti-correlated with
 $t_p$, the preceding sentence points towards a consistent picture in which
  the first peak is frequently harder and wider relative to the last peak.
 While, in the case of Class B bursts, more often than not first peaks tend
 to be relatively narrower, weaker and softer.
  However, flux and hardness ratio of peaks are strongly correlated in
 both the groups.

\end {section}
\begin {section}{Conclusion}
  Studying bursts with T90 less than 2 s, we observe that
 irrespective of hard-to-soft evolution or vice-versa, there exists a
  statistically
 significant correlation between number of subpulses and burst duration,
  indicating
  that if injection of energy from central engine takes place in the form of multiple
  spurts of wind then longer the activity
 of the engine, more is the number of shocked shells that give rise to pulses of
 gamma emission.

 If one were to accept a genuine duration-fluence correlation (see the
 discussion in Sec.3)
  then from the results listed in table 1 one
  may sketch a general picture as follows.
  When total amount of energy available is larger, it is natural
   for Lorentz factors of ejecta as well as fluence to be greater, providing
  a plausible explanation for the strong correlation observed between HR and
  fluence. Then, the T90-fluence correlation appears to indicate that when
  total energy involved is more in magnitude,
  instead
  of an instaneous conversion to gamma photons,
  it tends to be released and dissipated over a
 longer time-scale resulting in larger number of pulses as well as higher
  values of $\vert \tau_c \vert$.

 We find from tables 2 and 3 that bursts with lag $\tau_c < 0 $ show
 different behaviour from
 the ones with $\tau_c > 0$.  Among the Class A objects,  the height, count
 and hardness
 ratio of a pulse show strong trend of decreasing with time,
 which is more or less expected from progressive dissipation of energy.
 In contrast, the tendency of peak height, number of photons and
 HR of individual pulses to increase with peak position among Class B bursts
 seems to suggest that
 the conversion
 of available energy into stronger pulses consisting of harder gamma photons
 takes place during the later part
 of GRB activity.
 One may surmise that  the  central device becomes vigorously active
 towards the later part before
  getting switched off completely, possibly
  due to a delayed accretion of matter on to a compact object that makes up the
 engine.

  It appears to be significant that only $\sim $ one-fifth of the set of
 single peaked bursts
  happen to be of Class B type. A flare up of the central source in the later
 half is likely to give rise to additional pulses of radiation. Hence, in this picture, it
 is not surprising for CI to be often $>1$ in Class B category.

 Unlike the
  Class A bursts, the objects with soft-to-hard evolution hardly display
 any correlation between $\tau_c$ and fluence, CI and HR, respectively.
 Moreover, pulse parameters related to $tr$, $td$, $w$ etc. exhibit dissimilar
 trends for the two classes. On the whole, we may conclude that the sources
 demonstrating hard-to-soft spectral decay and those with opposite behaviour
 form two distinct categories of GRBs.
\end {section}

\begin{table}
\begin{center}
\begin{tabular}{|lllll|}\hline

X &  Y & N     & Correl. & Chance     \\
  &              &                     & Coeff. & Prob. \\
\hline
&&&&\\
$CI$     & $T_{90}$ & 156 & 0.59        & $4.03\times10^{-8}$ \\
$CI$  & Fluence & 149 & 0.43 & $6\times 10^{-8}$\\
$CI$  & HR & 133 & 0.49          & $4.2\times10^{-8}$ \\

&&&&\\
$\vert \tau_c \vert $& $T_{90}$ & 156 & 0.48 &$1.0\times10^{-8}$\\

&&&&\\
$T_{90}$ & Fluence& 149 & 0.43&$6\times10^{-8}$\\
&&&&\\
 Fluence & HR & 133 &  0.51 &$4.07\times10^{-8}$ \\

&&&&\\
&&&&\\

\hline
\end{tabular}
\vspace{0.2in}

\caption{Correlation coefficients and corresponding chance probabilities
 for the sample of short GRBs. $X$ and $Y$ are the burst parameters while $N$
 is the number of data points used in the correlation analysis}
\end{center}
\end{table}

\begin{table}
\begin{center}
\begin{tabular}{|llllllll|}\hline &&&&&&&\\
&&$\tau_c$&$\ \ \ >$&0 &$\tau_c$&$\ \ <$ & 0\\  &&&&&&&\\ \hline &&&&&&&\\

X & Y & N     & Correl. & Prob. & N & Correl. & Prob.     \\
&&&&&&&\\&&&&&&&\\
$ \tau_c $&CI& 40 & -0.016 &0.92& 116 &-0.20 &0.03\\
$\tau_c$&$T_{90}$ &40  &0.56 &.0002& 116 & -0.52&$4.25\times 10^{-8}$\\
$\tau_c$&Fluence & 37 & 0.09 &0.6 & 112 &-0.41&$7.9\times 10^{-6}$\\
$\tau_c$ & HR & 33 & -0.03&0.87& 100 &  -0.27&.006\\
&&&&&&&\\&&&&&&&\\
$T_{90}$&HR& 33 & 0.53 &.002 & 100 & 0.09 &0.2\\
&&&&&&&\\&&&&&&&\\
\hline
\end{tabular}
\vspace{0.2in}
\caption{Correlations among burst parameters X and Y, separately for positive
  and
  negative time lags $\tau_c$. N is the number of observed pairs of data.}
\end{center}
\end{table}

\begin{table}
\begin{center}
\begin{tabular}{|llllllll|}\hline &&&&&&&\\
&&$\tau_c$&$>$ &\ \ \ 0&$\tau_c$&$< $ &\ \ 0\\  &&&&&&&\\ \hline &&&&&&&\\

 X & Y     & Correl. & N  & Prob. & Correl. & N & Prob.     \\
&&&&&&&\\ \hline &&&&&&&\\
$t_p$ & $n_{ph}$ &0.23 &69&.06& -0.42&123&$1.4\times10^{-6}$\\
$t_p$ & $hr_{pk}$ &0.3 &69 & .012& -0.36& 123& $4.3\times10^{-5}$\\
&&&&&&&\\
\hline
&&&&&&&\\
$\tau_c$&$(hr_{pk})_{first}$&-0.45 & 20 & 0.05 &-0.35& 41 & 0.02\\
$\tau_c$&$w_{first}$&0.38 & 20 & 0.1 &-0.34& 41 & 0.03\\
$\tau_c$&$(r_{rd})_{first}$&-0.04 & 20 & 0.87 &0.29& 41 & 0.07\\
$\tau_c$&$ td_{first}/td_{last}$&0.34 & 20 & 0.007 &-0.045& 41 & 0.74\\
\hline &&&&&&&\\
$hr_{pk}$&$flx_{pk}$&0.21& 69 & 0.08&0.67& 123 & $4 \times 10^{-8}$\\
$(hr_{pk})_{first}$&$(r_{rd})_{first}$&0.45& 20 & 0.05&-0.16 & 41 & 0.32\\
$(hr_{pk})_{first}/(hr_{pk})_{last}$&$tr_{first}/tr_{last}$&0.43& 20 & 0.06&0.32& 41 & 0.04 \\
$ (hr_{pk})_{first}/(hr_{pk})_{last}$&$td_{first}/td_{last}$&0.45& 20 & 0.05&0.46&41 &0.002 \\
$(hr_{pk})_{first}/(hr_{pk})_{last}$&$w_{first}/w_{last}$&0.55& 20& 0.01&0.48& 41 &0.001 \\
&&&&&&&\\
\hline
\end{tabular}
\vspace{0.2in}
\caption{Correlation coefficients for 61  short bursts with well separated
 peaks.}
\end{center}
\end{table}
\begin{figure}

\vspace{-1.0 cm}
\hspace{-2.0 cm}
\makebox

{

\epsfig{file=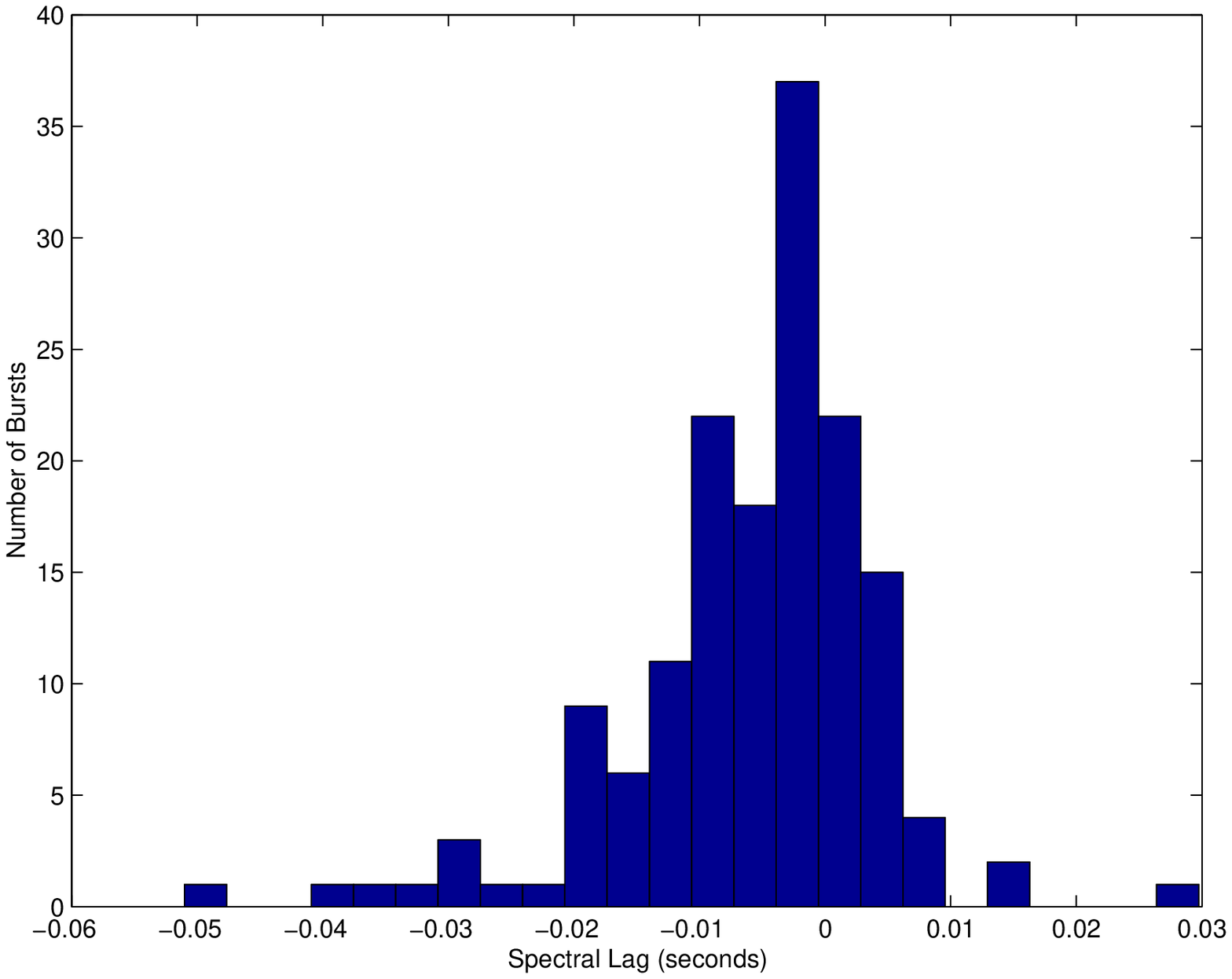,height=10.0cm,width=15.0cm} }
\caption{Distribution of Spectral Lag $\tau_c$.}

\end{figure}


\begin {thebibliography}{Reference}
\bibitem { } Kouveliotou,C.,Meegan,C.A.,Fishman,G.J.,Bhat,P.N.,Briggs,M.S.,Koshut,T.M.,
Paciesas,W.S. and Pendleton,G.N., Ap.J.413,L101 (1993)
\bibitem { }Mukherjee,S. et al., Ap.J.508,314 (1998)
\bibitem { }Panaitescu,A.,Kumar,P. and Narayan,R., astro-ph/0108132
\bibitem { }Norris,J.P.,Scargle,J.D. and Bonnell,J.T.,astro-ph/0105108
\bibitem { }Paciesas,W.S,Preece,R.D.,Briggs,M.S. and Mallozzi,R.S., astro-ph/0109053
\bibitem { }McBreen,S.,Quilligan,F.,McBreen,B.,Hanlon,L. and Watson,D., astro-ph/0112517
\bibitem { }Norris,J.P.,Marani,G. and Bonnell,J.,Ap.J.534,248 (2000)
\bibitem { }Bhat,P.N.,Fishman,G.J.,Meegan,C.A.,Wilson,R.B. and Paciesas,W.S.,AIP Conference Proceedings 307, Second Workshop on Gamma
  Ray Bursts, Huntsville, 1994,p-197
\bibitem { }ftp://cossc.gsfc.nasa.gov/batse
\bibitem { }Bhat,P.N.,Gupta,V. and Das Gupta,P.,5th Compton Symposium, AIP Conference Proceedings 510, 2000,p-538
\bibitem { }Band,D.L.,Ap.J.486,928 (1997) and the references there in.
\bibitem { }Norris,J.P. et al., Ap.J.459, 393 (1996)
\bibitem { }Quilligan,F. et al., astro-ph/0112515
\bibitem { }Bevington,P.R.,Data Reduction and Error Analysis for the Physical Sciences,
McGraw-Hill Book Company, New York, 1969.
\bibitem { }Rees,M.J. and Meszaros,P., Ap.J.430,L93 (1994)
\bibitem { }Fenimore,E.E. et al. Nature 366,40 (1993)
\bibitem { }Lee,T. and Petrosian,V.,Ap.J.474,37 (1997)
\bibitem { }Hakkila,J. et al., in AIP Conf.Proc.526, Gamma Ray Bursts,2000,p-48
\bibitem { }Hakkila,J. et al., in AIP Conf.Proc.526, Gamma Ray Bursts,2000,p-33
\bibitem { }Gupta,V., Das Gupta,P. and Bhat,P.N., in AIP Conf.Proc.526, Gamma Ray Bursts,    2000,p-215

\end {thebibliography}
\end{document}